%% file: conference_main.tex
\documentclass[runningheads]{llncs}

\usepackage{booktabs}
\usepackage{graphicx}
\usepackage{array}
\usepackage{pifont}
\usepackage{makecell}
\usepackage{adjustbox}
\usepackage{xcolor}
\usepackage{etoolbox}

\usepackage[T1]{fontenc}
\usepackage{graphicx}
\usepackage{todonotes}
\usepackage{amsmath}

\begin{document}

\title{ESL-former: Event Sparse Linear Transformer for FPGA-Accelerated Malware Detection}

\titlerunning{ESL-former for FPGA-Accelerated Malware Detection}



\author{Jun Yang\inst{1} \and Yumeng Cao\inst{1} \and Yingxuan Xiao\inst{1} \and Changhong Li\inst{2}\textsuperscript{*} \and Shangru Zhou\inst{3}}

\authorrunning{Yang et al.}
%

\institute{School of Computer Science and Engineering, Changsha University, China \and Electronics and Electrical Engineering, Trinity College Dublin, Ireland \and School of Mechanical and Electrical Engineering, Changsha University, China}

\maketitle
\input{sections/abstract}

\input{sections/sec1_introduction}
\input{sections/sec2_related_works}
\input{sections/sec3_methodology}

\input{sections/sec4_results}

\input{sections/sec5_conclusion}

\subsubsection*{Funding}
This work is supported in part by funds from the
Regional Joint Fund Project of Natural Science
Foundation of Hunan Province (2025JJ70016),
Outstanding Young Scholars Program of the Scientific
Research Projects of the Education Department of
Hunan Province (25B0813).

\bibliographystyle{unsrt}
\bibliography{references}

\end{document}

%% file: sections/abstract.tex
\begin{abstract}
Transformer based intrusion detection has shown strong potential for identifying unknown malware variants, but its deployment on resource constrained edge devices remains difficult due to the high computational and memory cost of self attention. This paper presents an event sparse Linear Transformer framework for malware detection on FPGA edge platforms. Instead of relying on an explicit attention matrix, the proposed method derives token and channel importance directly from the accumulated key value state of Linear Transformer. A contribution aware scoring function is introduced to measure the perturbation of each token or kernel channel to the linear attention state, allowing redundant tokens and low contribution channels to be removed without constructing quadratic attention maps. To translate sparsity into practical hardware acceleration, the proposed design further replaces conventional binary mask based execution with a Top $K$ active token buffer and a hardware compaction unit. The selected active tokens and channels are reorganized into dense fixed length data streams, enabling downstream processing elements to execute regular multiply accumulate operations with reduced redundant memory access and control overhead. This design is suitable for low power FPGA based malware detection, where both latency and on chip resource usage are critical. Experimental evaluation demonstrates that the proposed sparse Linear Transformer can reduce hardware cost while preserving malware classification accuracy.
\end{abstract}

\keywords{Malware detection \and Linear Transformer \and Event sparsity \and FPGA acceleration \and Edge intelligence}

%% file: sections/sec1_introduction.tex
\section{Introduction}
\label{sec:introduction}
The rapid evolution of malware and network attacks has placed increasing pressure on traditional rule-based and signature-based malware detection systems.
Rule-based engines remain important in high-speed network security systems, which offload Snort-compatible malware prevention workloads to FPGA hardware for 100~Gbps packet processing~\cite{ref1}.
However, such systems are primarily driven by predefined rules and signatures, which limits their adaptability when facing zero-day malware, obfuscated samples, and highly variant attack behaviours.
This limitation has motivated the use of learning-based detection methods for malware and malware detection.

Deep learning-based detection methods can automatically learn discriminative semantic patterns from traffic features, binary code, or malware image representations.
Transformer models provide stronger contextual modelling capability through self-attention, making them attractive for detecting complex malware patterns and highly variant malicious behaviours.
Despite their modelling capability, Transformer models impose substantial computation and memory overhead. Standard self-attention explicitly constructs a token-to-token attention matrix, whose computational and memory complexity grows quadratically with the sequence length~\cite{ref6,ref9,ref10}.

FPGAs are an ideal platform for edge IDS deployment due to their flexibility and high-speed computing capabilities.
Recent FPGA Transformer accelerators have attempted to reduce this cost using window attention~\cite{ref6}, complete FPGA mapping flows for large language model inference~\cite{ref7}, low-complexity encoder-only Transformer designs~\cite{ref8}, and linear attention acceleration for vision Transformer workloads~\cite{ref10}. 
These studies demonstrate the feasibility of efficient Transformer deployment on FPGA platforms, but resource-constrained malware detection on small edge FPGAs still requires further optimisation.

Linear Transformer offers an attractive alternative by replacing explicit softmax attention with a kernel-based formulation. Instead of materialising the full attention matrix, it accumulates key-value information into a compact global state, thereby reducing the dominant attention complexity from quadratic to linear with respect to the sequence length~\cite{ref9,ref10}.
This property is particularly suitable for edge FPGA deployment because it reduces attention memory pressure and avoids storing dense attention maps. However, the linear attention formulation also changes the basis for sparsification.
Many existing sparse Transformer methods are designed around standard attention, where token importance can be estimated from attention maps, dynamic masks, or structured sparse patterns~\cite{ref6,ref11,ref13}.
These assumptions do not directly apply to Linear Transformer, since the attention matrix is not constructed during inference.

Therefore, an effective sparse Linear Transformer for FPGA-based malware detection requires two key components.
First, the sparsity metric should be mathematically aligned with the key-value state update mechanism of Linear Transformer, rather than relying on an explicit attention matrix.
Second, the sparse execution scheme should translate algorithmic sparsity into actual hardware savings, including fewer loop iterations, reduced BRAM accesses, and fewer multiply-accumulate operations.
Prior sparse Transformer accelerators have explored dynamic sparse attention~\cite{ref11} and full N: M sparsity for FPGA-based Transformer architectures~\cite{ref13}.
However, mask-based or irregular sparse execution can still introduce additional control overhead on FPGA, such as mask loading, conditional checking, sparse index handling, and non-contiguous memory access.
These overheads are particularly harmful for small edge FPGAs.

To address these challenges, this paper proposes ESL-former, an Event Sparse Linear Transformer framework for FPGA-accelerated malware detection.
The framework first adopts a lightweight Linear Transformer backbone to avoid explicit quadratic attention computation.
It then introduces an event sparsity mechanism inspired by the event-driven sparsity concept in spiking Transformer acceleration~\cite{ref14}.
The proposed mechanism derives token and channel importance directly from the accumulated key-value state of the Linear Transformer.
Rather than using an external prediction module or an explicit attention map, ESL-former measures how strongly each token or kernel channel contributes to the runtime linear attention state.
Low-contribution tokens and channels are removed after kernel feature mapping and before the key-value state update, thereby reducing redundant computation in the actual attention dataflow.


The main contributions of this paper are summarised as follows.

\begin{itemize}
\item We propose ESL-former, a lightweight event-sparse Linear Transformer framework for FPGA-accelerated malware detection on resource-constrained edge platforms.

\item We introduce a contribution-aware token and channel scoring method that is directly derived from the key-value state accumulation of Linear Transformer.

\item We design a hardware-oriented Top-K sparse execution scheme that replaces binary mask-based sparsity with active token buffering and channel compaction.

\item We implement and evaluate ESL-former on the PYNQ-Z2 FPGA platform for IMG\_DS malware image detection. Experimental results show that ESL-former achieves up to 41.56$\times$ end-to-end speedup over ViT4Mal while maintaining competitive detection performance. 
\end{itemize}

The remainder of this paper is organised as follows.
Section~\ref{sec:methodology} describes the proposed event-sparse Linear Transformer and its hardware-oriented sparse execution scheme.
Section~\ref{sec:results} presents the experimental results on quantization, sparse execution,on-board hardware performance,and malware detection.
Finally, Section~\ref{sec:conclusion} concludes the paper.


%% file: sections/sec2_related_works.tex
\section{Related Works}\label{sec:related_works}

\subsection{FPGA Accelerated Malware Detection}

With the continuous growth of malware targeting Internet of Things and industrial systems, FPGA-based security acceleration has attracted increasing attention.
FPGAs can execute customised packet processing, feature extraction, and machine learning inference pipelines with high parallelism and low power consumption, making them suitable for real-time edge security applications. Pigasus~\cite{ref1} implements a high-speed intrusion prevention system using FPGA acceleration for Snort rule matching.
It offloads most packet inspection operations to the FPGA and leaves only a small portion of complex cases to the CPU, thereby improving throughput and reducing host processing overhead.

In addition to rule-based acceleration, several studies have mapped machine learning models onto FPGAs for network intrusion detection, malicious traffic classification, and IoT attack detection. Representative examples include FPGA acceleration for graph neural networks~\cite{ref2}, lightweight LSTM computation~\cite{ref3}, and XGBoost-based ransomware detection~\cite{ref4}. OmBNNet~\cite{ref5} further explores a resource-efficient binarised neural network for obfuscated Android malware detection on FPGA. MalViT~\cite{zou2025jpeg} reduces the computational cost of malware image classification through lightweight model design. However, these approaches mainly focus on reducing model complexity on general computing platforms and do not consider FPGA-specific resource constraints and hardware execution efficiency.These methods demonstrate that FPGA-based machine learning inference can support low-power security analytics.
However, many of these models have limited capability in modelling long-range dependencies and global contextual relations, which are important for detecting highly variable malware samples and complex multi-stage attacks.Hardware-assisted malware detection has also been investigated from different perspectives. Patel et al.~\cite{ref15} evaluated FPGA implementations of machine-learning-based malware detectors in terms of latency, area, and power,while DroidHunter~\cite{ref16} explored vision-based Android malware detection using opcode-derived image representations.

\subsection{Efficient Transformer Models and FPGA Deployment}

Transformer models have been introduced into security related tasks because self attention can capture long range dependency and global contextual information from sequential or image based malware representations. A recent systematization by Kunwar et al.~\cite{ref17} further summarizes the growing use of Transformer architectures for malware detection, classification, and feature representation.However, standard self attention requires the computation of an explicit token to token attention matrix, resulting in high memory consumption and quadratic complexity with respect to the sequence length. This cost is particularly problematic on FPGA edge platforms, where BRAM, DSP, and off chip memory bandwidth are limited.

Recent studies have attempted to reduce the cost of Transformer deployment on FPGAs through attention approximation, dataflow optimization, quantisation, and specialized hardware mapping. SWAT~\cite{ref6} proposes a sliding window attention based Transformer accelerator, which constrains the attention range of tokens and improves latency and energy efficiency on FPGA. FlightLLM~\cite{ref7} targets large language model inference and uses FPGA DSP48 resources to construct configurable sparse DSP chains, together with a length adaptive compilation method for reducing instruction storage overhead. SATNet~\cite{ref8} proposes a low complexity encoder only Transformer design using second order approximation to reduce the cost of nonlinear operations. cosFormer~\cite{ref9} rethinks softmax attention and proposes a linear attention alternative based on non negative feature mapping and cosine based reweighting. FastViT~\cite{ref10} develops a real time linear attention accelerator for dense vision prediction tasks using efficient multiplication, mixed precision quantisation, and optimized dataflow.

These works indicate that efficient Transformer inference on FPGA is feasible, but most of them are designed for relatively high performance FPGA platforms or general vision and language workloads. For small edge FPGAs used in security applications, the memory and control overhead of Transformer inference still remains a key bottleneck. In particular, malware detection requires a design that can jointly consider detection accuracy, latency, on chip memory usage, and the practical executability of sparse attention on limited FPGA resources.

\subsection{Sparse Transformer Acceleration}

Sparsity has become an important direction for reducing Transformer computation. Existing sparse Transformer methods mainly exploit sparsity from attention patterns, token redundancy, or weight matrices. SWAT~\cite{ref6} reduces attention computation through window-constrained sparse attention. FlightLLM~\cite{ref7} explores weight sparsity and sparse DSP mapping for efficient large language model inference. Other FPGA-based designs introduce dynamic sparse attention, where low-importance tokens are removed by prediction or masking mechanisms~\cite{ref11}. ASAP~\cite{ref12} adopts alternating sparsity and asymmetric precision to process dense and sparse weights efficiently, while FNM-Trans~\cite{ref13} uses N sparsity to optimise QKV generation, attention computation, and feed-forward network modules.

Dynamic token sparsification has been extensively investigated for reducing
the input-dependent computation of Vision Transformers.
DynamicViT~\cite{rao2021dynamicvit} inserts lightweight prediction modules
into multiple Transformer stages and progressively suppresses redundant
tokens according to the current input.
EViT~\cite{liang2022not} uses class-token attention to identify attentive
patches and fuses inattentive tokens so that part of their information is
preserved during subsequent layers.
Although these methods provide favourable accuracy--computation trade-offs,
their token scores are coupled with standard multi-head self-attention,
learned prediction modules, or class-token interactions.
Consequently, they cannot be directly transferred to a Linear Transformer
whose attention is represented by an accumulated key--value state without
constructing an explicit attention map.
In contrast, ESL-former evaluates each token according to the magnitude of
its direct update to the kernelised key--value state, aligning token selection
with the actual forward computation of linear attention.

Although these methods reduce theoretical computation, their hardware efficiency depends heavily on how sparsity is represented and executed. Dynamic masks, irregular sparse indices, and structured compression often introduce additional control logic, memory storage, and scheduling overhead. On FPGA, a sparse algorithm may fail to provide proportional latency reduction if the accelerator still traverses the original dense loop structure or repeatedly accesses irregular memory locations. This issue is especially important for edge FPGAs, where the cost of sparse control and buffering can offset the benefit of reduced arithmetic operations.

Another limitation is that many sparse Transformer methods are designed around standard softmax attention. Their importance scores are often obtained from explicit attention maps, learned token predictors, or class token attention. These mechanisms are not directly suitable for Linear Transformer, where attention is represented through a compact key-value state and no explicit attention matrix is materialised~\cite{katharopoulos2020transformers}. Therefore, sparse Linear Transformer acceleration requires a scoring strategy coupled with the state accumulation mechanism itself.

\subsection{Position of This Work}

Different from prior sparse Transformer accelerators that mainly exploit sequence-level attention sparsity or weight-level numerical sparsity, this work focuses on event-sparse execution for Linear Transformer-based malware detection on a resource constrained FPGA. ESL-former derives token and channel importance from the key-value state update of Linear Transformer, which allows redundant tokens and low-contribution kernel channels to be removed without constructing quadratic attention maps. This makes the sparsification process consistent with the actual runtime computation of Linear Transformer.

Furthermore, ESL-former emphasizes hardware executable sparsity. Instead of only applying binary masks to suppress inactive tokens, the proposed design uses Top K active token buffering and channel compaction to reorganise sparse activations into dense fixed-length streams. As a result, downstream FPGA processing elements can execute regular dense multiply-accumulate operations with fewer effective loop iterations and reduced memory traffic. This hardware-oriented sparse dataflow distinguishes ESL-former from methods that reduce arithmetic operations at the algorithmic level but retain a largely dense or irregular execution path on FPGA.

%% file: sections/sec3_methodology.tex
\section{Methodology}\label{sec:methodology}
When an unknown executable infiltrates a computer system, it must be detected as early as possible before causing further damage.
This requirement calls for the rapid extraction of discriminative behavioural features that can effectively distinguish benign activities from malicious ones.
However, edge devices suffer from limited computational and memory resources; achieving low-latency malware detection on resource-constrained devices remains highly challenging.

We propose a Linear Transformer-based detection framework that performs
dynamic sparsification along both the token and kernel-feature channel dimensions of the query and key representations
after kernel feature mapping. The proposed method investigates how
different quantisation schemes and sparsity levels affect the efficiency and detection
performance of malware classification on resource-constrained edge systems.
Figure 1 illustrates the overall architecture of the proposed framework.
The original IMG\_DS samples are \(299\times299\) RGB JPEG images. In the preprocessing pipeline used for the reported experiments, each image is converted to grayscale and uniformly resized to \(32\times32\) pixels using bilinear interpolation.
Each resized image is then partitioned into non-overlapping patches of size
\(8\times8\), producing a \(4\times4\) patch grid and an encoder sequence length of \(N=16\). Each patch is flattened into a 64-dimensional token and projected to the hidden/kernel-feature width \(d=16\) before being fed into the Transformer encoder.
The encoder incorporates our proposed event-sparse linear self-attention mechanism,
along with multilayer perceptron, layer normalization, and residual connections,
to capture global feature representations of the input images. The decoder subsequently
processes the output from the final encoder layer and performs binary
classification into malware or benign categories via an MLP followed by a Softmax
activation function.

In the following sections, we will introduce the proposed sparse linear transformer, its sparse selection algorithm, and the hardware implementation architecture of the event sparsity mechanism.

\begin{figure}[!ht]
    \centering
    \includegraphics[width=1\linewidth]{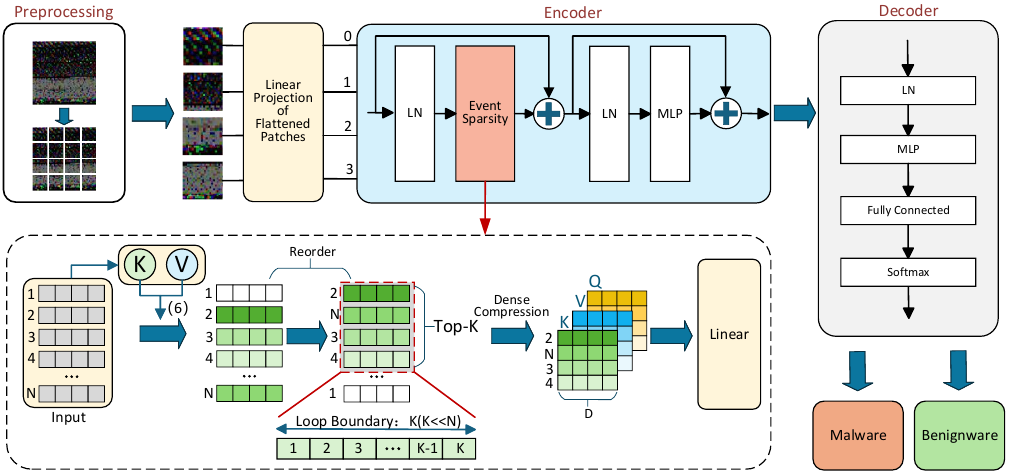}
    \caption{ESL-former overview: Event Sparse Linear Transformer Framework for Malware Detection.}
    \label{fig:framework}
\end{figure}

\subsection{Linear Transformer}

Given the query, key, and value matrices \(Q\), \(K\), and \(V\), standard attention is computed as
\begin{equation}\label{eq:standard_attention}
    \mathrm{Attn}(Q,K,V) = \mathrm{softmax} \left( \frac{QK^\top}{\sqrt{d}} \right)V, 
\end{equation}
where \(N\) denotes the sequence length and \(d\) denotes the hidden dimension. As shown in Eq.~\eqref{eq:standard_attention}, the computation requires an explicit \(N \times N\) attention matrix. This leads to substantial memory access and computational bottlenecks, especially when processing long sequences on resource constrained edge platforms.

Linear Transformer mitigates this issue by approximating the attention operation through a kernel based factorization. In this work, we adopt a non negative kernel feature mapping \(\phi(\cdot)\), implemented as \(\mathrm{elu}(\cdot)+1\). After mapping \(Q\) and \(K\) into the kernel feature space, the key value information can be accumulated into a compact global state:
\begin{equation} \label{eq:linear_state}
    S = \Phi(K)^\top V, 
\end{equation}
where \(\Phi(K)=\phi(K)\). When the value dimension is equal to the hidden dimension, the state \(S\) has size \(d \times d\). This compact state can be reused by all query tokens, avoiding the need to explicitly construct the full attention matrix.

Compared with the standard Transformer, Linear Transformer reduces the dominant computational cost from $\mathcal{O}(N^2 d)$ to $\mathcal{O}(N d^2)$. This linear formulation avoids explicitly materializing the \(N\times N\) attention matrix and is therefore attractive for resource-constrained FPGA deployment.

\subsection{Channel Contribution Importance Evaluation}


However, the absence of an explicitly constructed attention matrix makes traditional token importance estimation methods difficult to apply directly. Many existing pruning methods rely on attention weights, an $N \times N$ attention map, or the class token to aggregate global information—assumptions that do not directly hold for linear Transformers. Consequently, it is necessary to derive a token importance metric directly from the internal states of the Linear Transformer to enable effective token-level dynamic sparsification.


Based on Eq.~\eqref{eq:linear_state}, the importance score of the \(c\)-th kernel channel is defined as
\begin{equation} \label{eq:channel_score}
    \alpha_c = \left\| S_{c,:} \right\|_1,
\end{equation}
where \(S_{c,:}\) denotes the \(c\)-th row of the compact state. It quantifies how strongly each channel contributes to the global key-value representation used by subsequent query tokens.

After obtaining the channel scores, we apply a Top-\(k\) sparsification strategy to select the most informative channels. Specifically, only the channels with the highest \(k\%\) contribution scores are retained:
\begin{equation}\label{eq:topk_channel}
    \mathcal{C}_{k} = \operatorname{TopK} \left( \{\alpha_c\}_{c=1}^{d}, \left\lceil k d \right\rceil \right), 
\end{equation}
where \(d\) is the number of kernel channels and \(\mathcal{C}_{k}\) denotes the selected active channel set. This scheme preserves dominant channels while removing low-contribution ones from the Q/K kernel feature space, enabling Linear Transformer to reduce redundant computation and memory access while maintaining the channels most relevant to the runtime attention state.


Building upon the pruned channel space defined by $\mathcal{C}_{k}$, we further evaluate token importance. The core runtime state of the Linear Transformer, restricted to these active channels, is accumulated from key features and value representations. Specifically, the compact key-value state \(M\) can be written as
\begin{equation}\label{eq:linear_kv_state}
    M = \sum_{i=1}^{N} \phi(k_i) v_i^\top,
\end{equation}
where \(\phi(k_i)\) denotes the kernel-mapped key feature of the \(i\)-th token projected onto the active channels $\mathcal{C}_{k}$, and \(v_i\) denotes its corresponding value vector. Each token contributes an outer-product update to the global linear attention state.

If the \(i\)-th token is removed, the accumulated state decreases by its corresponding update term. Therefore, the direct contribution of this token to the current linear attention state can be quantified by measuring the magnitude of this state update. In this work, we define the importance score of the \(i\)-th token as the Frobenius norm of its contribution to the key-value state:
\begin{equation} \label{eq:token_score}
    s_i = \left\| \phi(k_i) v_i^\top \right\|_{F}.
\end{equation}


It measures the direct perturbation imposed on the Linear Transformer key-value state when the \(i\)-th token is removed. Compared with DynamicViT, which relies on an additional MLP predictor to indirectly learn proxy importance scores, our proposed score is directly derived from the forward computation graph. This makes it more aligned with the actual runtime state. Furthermore, explicitly constructing the outer-product matrix in Eq.~\eqref{eq:token_score} would introduce unnecessary overhead. Since $\phi(k_i)$ and $v_i$ are vectors, the Frobenius norm can be decomposed as:
\begin{equation} \label{eq:frobenius_decomposition}
    \left\| \phi(k_i)v_i^\top \right\|_{F} = \left\| \phi(k_i) \right\|_{2} \left\| v_i \right\|_{2}. 
\end{equation}



Based on this, the token score can be computed without materializing the \(d \times d\) outer-product matrix. In practice, for Top-\(k\) based sparsification, we use the equivalent squared score:
\begin{equation}\label{eq:squared_token_score}
    \tilde{s}_i = \left\| \phi(k_i) \right\|_{2}^{2} \left\| v_i \right\|_{2}^{2},
\end{equation}
which preserves the ranking while avoiding square-root operations. As shown in Eq.~\eqref{eq:squared_token_score},the computation requires only two vector-wise squared sums and one scalar multiplication. For two length-\(d\) vectors, the hardware cost is approximately \(2d+1\) arithmetic operations, with only \(\mathcal{O}(d)\) temporary storage. 


\subsection{Token Sparsification and Hardware Compression}
\begin{figure}[!ht]
    \centering
    \includegraphics[width=1\linewidth]{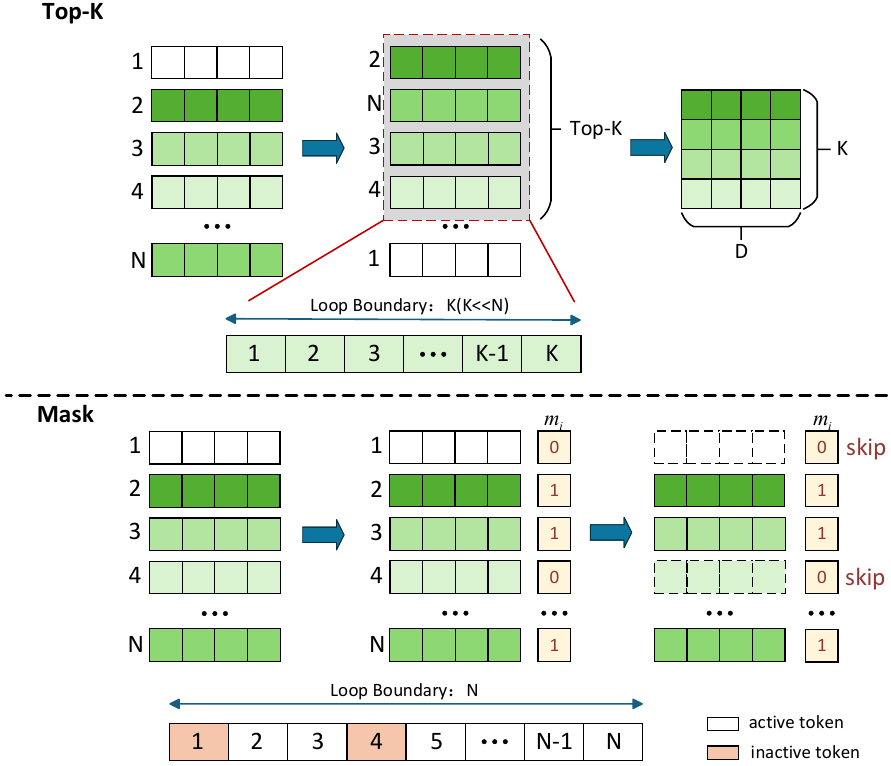}
    \caption{Differences between Top-K sparsity and mask sparsity.}
    \label{fig:topk_schematic}
\end{figure}


To distinguish active tokens from inactive ones, a straightforward implementation is to construct a binary token mask. Let $m_i$ denote the mask value for the $i$-th token, where $m_i=1$ indicates the token is retained, and $m_i=0$ indicates it is suppressed. Based on this, the masked state update of the Linear Transformer is formulated as:
\begin{equation} \label{eq:masked_state_update}
    M_{\mathrm{mask}} = \sum_{i=1}^{N} m_i \, \phi(k_i)v_i^\top.
\end{equation}


Although mask-based sparsification mathematically removes the contribution of low-score tokens, it does not necessarily translate into proportional hardware savings on FPGAs. In a direct implementation, the hardware accelerator typically follows the original loop structure over the full token sequence ($N$). Consequently, the Query, Key, and Value (Q, K, V) vectors may still need to be generated for all tokens before the mask is applied. Even if inactive tokens are skipped within the loop, they still occupy loop iterations and incur overhead from mask loading, conditional branching, and irregular memory access. As a result, mask-based sparsification primarily reduces effective arithmetic operations, while the hardware execution path and control overhead remain close to the dense implementation.


To convert token sparsity into practical latency and BRAM-access reduction, we adopt a Top-$K$ based sparsification strategy. As shown in Figure \ref{fig:topk_schematic}, Instead of iterating over all tokens with a binary mask, we select the most informative tokens and  performs reordering subsequently, then compact them into a dense buffer, so that the subsequent sparse KV-state update and backend computation are performed on the compacted active-token representation.
Given the token contribution scores $s_i$ derived from Eq. \ref{eq:token_score}, the active token set $\mathcal{A}_{K}$ is selected as:
\begin{equation} \label{eq:topk_active_tokens}
    \mathcal{A}_{k} = \operatorname{TopK} \left( \{s_i\}_{i=1}^{N}, K \right),
\end{equation}
where $K$ is the predefined number of retained tokens, $D$ is the predefined number of retained Q/K kernel-feature channels. The selected tokens are then compacted into an active-token buffer 
$X_{\mathrm{active}} = \{x_i \mid i \in \mathcal{A}_{K}\}$. Consequently, the subsequent sparse KV-state update and backend computation are performed on these $K$ selected tokens, while contribution scoring and Top-K selection are still performed over the original $N$ candidate tokens.The key-value state and normalization term are updated as:

\begin{equation}
    M_{\mathrm{top}K} = \sum_{i \in \mathcal{A}_{K}} \phi(k_i)v_i^\top, \quad z_{\mathrm{top}K} = \sum_{i \in \mathcal{A}_{K}} \phi(k_i).
\end{equation}

The final attention output for a query $q_j$ is then computed as:
\begin{equation}
    y_j = \frac{\phi(q_j)^\top M_{\mathrm{top}K}}{\phi(q_j)^\top z_{\mathrm{top}K} + \epsilon}.
\end{equation}

This scheme shortens the effective loop bound from $N$ to $K$, directly reducing redundant loop iterations and memory traffic.

To support this dynamic sparsity on FPGA, we design a specialized hardware compaction unit. This unit evaluates the validity of each channel and token on the fly based on the calculated contribution scores. 
It first preserves the selected $D$ Q/K kernel-feature channels and then employs a multi-stage selector network to 
gather the $K$ selected token entries in a fixed order and concatenate them tightly. 


This process reconstructs the originally sparse and irregular representation into a compact dense $K\times D$ data block. 
The compacted output is forwarded as a dense submatrix to the downstream Processing Elements (PEs)
, so that the sparse attention backend operates on the $K$ selected tokens instead of traversing the full $N$-token loop with repeated mask checks.This ensures that the PE array can execute continuous dense multiply-accumulate operations without stalling for irregular memory accesses or checking sparse masks. By transforming mathematical sparsity into a hardware-friendly dense execution pattern, the proposed design significantly improves dataflow regularity and reduces DSP usage or active cycles.

%% file: sections/sec4_results.tex
\section{Results and Discussion}\label{sec:results}

\subsection{Experimental Setup}

We conducted experiments on the IMG\_DS malware image dataset to evaluate the effectiveness of the proposed algorithm. IMG\_DS is an IoT malware image dataset consisting of benignware and malware samples represented as images~\cite{ravi2023vit4mal}. The dataset contains 14,733 malware samples and 2,486 benign samples. In our framework, each malware image is divided into multiple patches, which are then embedded as input tokens for the Linear Transformer. The proposed framework considers sparsity optimization from both token and channel dimensions. Specifically, $N=16$ and $d=16$ denote the original sequence length and kernel-feature width, respectively.
$K$ denotes the number of retained active tokens after contribution-aware token selection, while 
$D$ represents the number of retained 
Q/K kernel-feature channels after channel contribution evaluation. 
We evaluate K16/D16, K14/D13, K13/D10, and K11/D8, with sparsity ratios defined as $1-K/N$ and $1-D/d$, respectively. The retention budgets are fixed, while the selected token and channel identities are dynamically determined for each input.
First, we evaluate different quantization configurations
, followed by a comparison between Mask Sparse and Event Sparse.
Finally, system-level deployment evaluation is conducted on the PYNQ-Z2 platform.


\subsection{Case Study 1: Sparse Unit Ablation}

This experiment compares the FPGA execution efficiency of Top-K based Event Sparse and binary mask-based Mask Sparse. To this end, we first evaluate three quantization configurations, Q32, Q16, and Q8, on the baseline Linear Transformer model, with the goal of determining a suitable deployment bit width for the subsequent sparsity ablation experiments. As shown in Table I, Q32 delivers high accuracy, but suffers from relatively large kernel latency. In contrast, Q16 introduces only a negligible accuracy degradation of less than 1 percentage point compared with Q32, while reducing the kernel latency from 3166.3~$\mu$s to 1146.1~$\mu$s.
Meanwhile, Q16 also reduces LUT and DSP resource utilization. Although Q8 further reduces the latency, its accuracy decreases to 93.00\%, indicating a larger performance degradation compared with Q16. Therefore, Q16 is selected as the deployment precision for subsequent token-channel sparse experiments.

\begin{table}[!ht]
\centering
\small
\setlength{\tabcolsep}{4pt}
\caption{Performance and Resource Comparison under Different Quantization Settings.}
\label{tab:sparse_core}
\begin{adjustbox}{max width=\columnwidth}
\begin{tabular}{l c c c c c c}
\toprule
\textbf{Quant} & 
\textbf{Acc (\%)} & 
\textbf{F1 (\%)} & 
\textbf{Kernel Latency ($\mu$s)} & 
\textbf{LUT}  &
\textbf{DSP}  &
\textbf{BRAM} \\
\midrule
Q32  & 95.00 & 97.11 & 3166.3 & 23431 & 184 & 9.0 \\
Q16 & 94.13 & 96.53 & 1146.1 & 13701 & 149 & 9.5 \\
Q8 & 93.00 & 95.91 & 891.3 & 14019 & 124 & 15.0 \\
\bottomrule
\end{tabular}
\end{adjustbox}
\end{table}

\begin{figure}[!ht]
    \centering
    \includegraphics[width=1\linewidth]{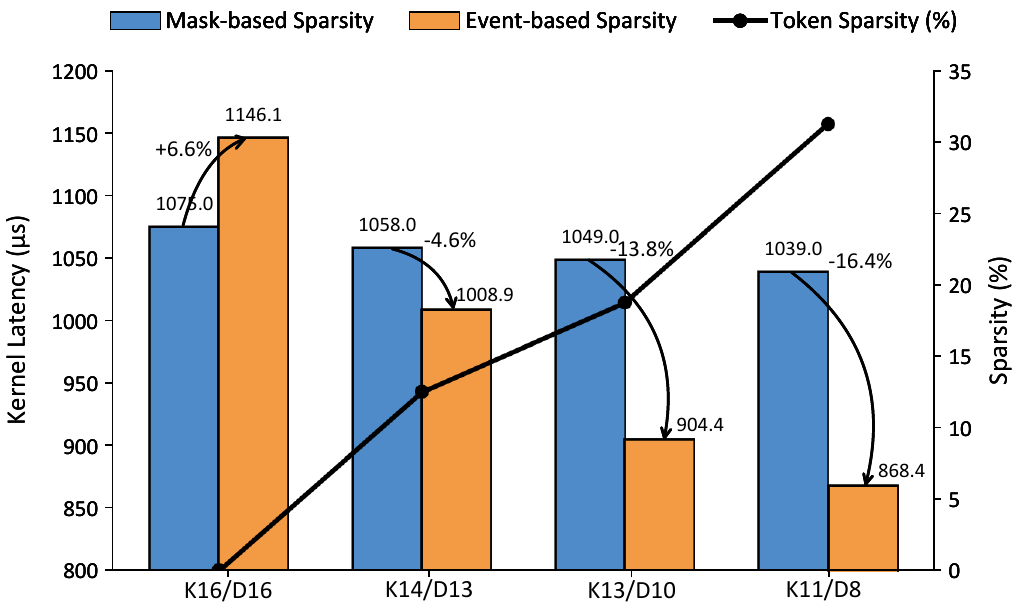}
    \caption{Kernel latency comparison between mask sparsity  and event sparsity  on PYNQ-Z2.}
    \label{fig:Event-MaskSparse}
\end{figure}

We first conducted a performance comparison between event-based sparsity and mask-based sparsity. In our experimental setup, we kept the Linear Transformer model, quantization bit width, sparsity ratio, and token importance scoring method identical across both approaches. Mask-based sparsity introduces a binary 0/1 indicator along the sequence dimension to determine whether each token participates in the state update. The results are shown in Fig.~\ref{fig:Event-MaskSparse}. The values following K on the x-axis indicate the number of retained tokens, where K16/D16 represents no token sparsity, K14/D13 corresponds to approximately 12\% token sparsity, K13/D10 to about 20\%, and K11/D8 to roughly 30\%.The corresponding D values indicate the number of retained feature channels selected by the contribution-aware channel evaluation mechanism. All data points are measured on the PYNQ-Z2 board.

As shown in Fig.~\ref{fig:Event-MaskSparse}, since mask-based sparsity merely masks computations within the loop without actually reducing the number of hardware iterations, the latency reduction across different sparsity levels is marginal. When the sparsity increases from K16/D16 to K11/D8, the latency decreases by only 36 $\mu$s (approximately 3\%). In contrast, Event Sparse achieves significant latency improvements by actively selecting effective tokens and optimizing the channel dimension through contribution-aware compression, which reduces redundant computation. Compared to the mask-based sparsity baseline, event-based sparsity incurs a 6.6\% latency overhead in the K16/D16 full-retention configuration due to the extra data reorganization cost. However, as the number of effective tokens decreases, its performance advantage becomes increasingly prominent, achieving kernel latency reductions of 4.6\%, 13.8\%, and 16.4\% under K14/D13, K13/D10, and K11/D8, respectively. These results demonstrate that by compressing active tokens so that subsequent KV updates process only valid tokens, event-based sparsity effectively translates sparsity into actual kernel acceleration under effective sparsity configurations.


\subsection{Case Study 2: On-Board Hardware Performance Evaluation}
This experiment aims to verify the practical hardware-level contribution of the proposed event-sparse FPGA design. Unlike mask-based sparsity, which only applies masking at the algorithmic level, this work maps sparsity directly onto the FPGA execution path through active token selection and contribution-aware channel compression, preventing low-contribution tokens and channels from entering subsequent computation processes. Therefore, this experiment focuses on whether event sparsity can effectively reduce hardware execution loops, on-chip memory accesses, and multiply-accumulate computation overhead. To demonstrate that the proposed design is not limited to algorithm-level sparsity, we conduct verification at three levels: detection performance, FPGA resource utilization, and on-board end-to-end latency on the PYNQ-Z2 platform.

\begin{figure}[!ht]
    \centering
    \includegraphics[width=1\linewidth]{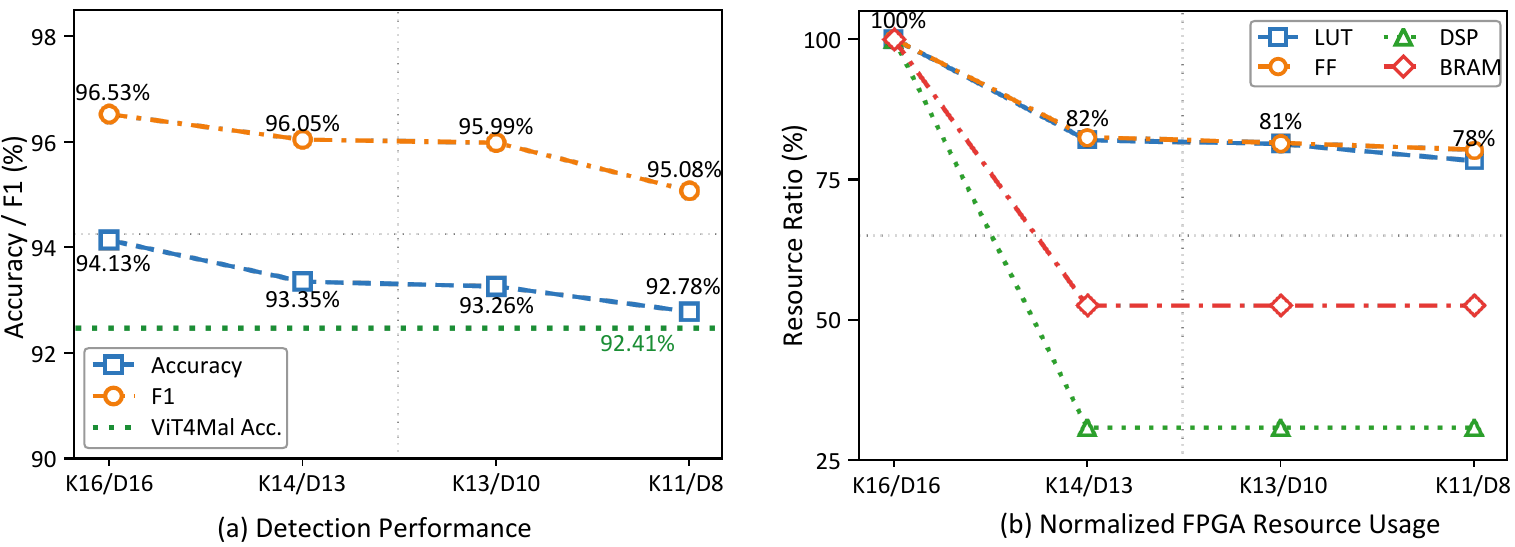}
    \caption{Detection performance and FPGA resource usage under different event-sparse configurations.}
    \label{fig:2}
\end{figure}

First, we compare the detection performance of different Token-Channel sparse configurations, as shown in Fig.~\ref{fig:2}(a). It can be observed that the detection performance remains generally stable as the sparse configuration changes from K16/D16 to K11/D8. The Accuracy and F1 of K16/D16 are 94.13\% and 96.53\%, respectively. K14/D13 and K13/D10 achieve Accuracy values of 93.35\% and 93.26\%, with F1 scores of 96.05\% and 95.99\%, respectively. When the compression level is further increased to K11/D8, the Accuracy and F1 decrease to 92.78\% and 95.08\%, respectively. These results indicate that Event Sparse can maintain stable detection performance under the K16/D16, K14/D13, and K13/D10 configurations, while K11/D8 reflects the performance degradation caused by a higher compression level.

Second, we analyze the FPGA resource variation introduced by event sparsity, as shown in Fig.~\ref{fig:2}(b). Fig.~\ref{fig:2}(b) presents the normalized FPGA resource usage under different Token-Channel sparse configurations, where K16/D16 is used as the reference configuration. The reported resource values correspond to fixed K/D hardware configurations, while runtime scoring determines which tokens and channels are retained.As the token number and channel dimension are further reduced, the LUT and FF utilization continuously decrease. The LUT utilization of K14/D13, K13/D10, and K11/D8 decreases to 0.82, 0.81, and 0.78, respectively, while the FF utilization decreases to 0.82, 0.81, and 0.80. The DSP and BRAM usage remains relatively stable across different event-sparse configurations, with the BRAM usage maintained at approximately 0.53. These results indicate that event sparsity does not achieve acceleration by replicating a large number of computational units. Instead, while reducing the number of effective tokens participating in subsequent computation and compressing the feature channel dimension, it also reduces the major logic and memory resource usage.

\begin{table}[!ht]
\centering
\footnotesize
\setlength{\tabcolsep}{5pt}
\renewcommand{\arraystretch}{1.15}
\caption{PYNQ-Z2 end-to-end latency breakdown ($\mu$s).}
\label{tab:e2e_breakdown}
\begin{tabular}{l c c c c}
\toprule
\textbf{Stage} & \textbf{K16/D16} & \textbf{K14/D13} &\textbf{K13/D10} & \textbf{K11/D8} \\
\midrule
Input Preparation & 24193.6 & 20032.9 & 18462.0 & 17698.7 \\
PS$\to$PL Transfer & 355.0 & 355.0 & 357.7 & 352.1 \\
\textbf{PL Kernel} & \textbf{1146.1} & \textbf{1008.9} & \textbf{904.4}& \textbf{868.4}\\
PL$\to$PS Transfer & 325.2 & 323.5 & 327.8 & 324.2 \\
Post-processing & 246.8 & 238.5 & 238.4 & 238.4 \\
Other & 5.3 & 5.3 & 5.2 & 5.1 \\
\midrule
\textbf{Total E2E} & \textbf{26272.0} & \textbf{21964.1} & \textbf{20295.5}& \textbf{19486.9}\\
\bottomrule
\end{tabular}
\end{table}

Finally, we conducted real-board end-to-end measurements on the PYNQ-Z2 platform and decomposed the inference latency, as shown in Table~\ref{tab:e2e_breakdown}. The complete inference pipeline consists of input preparation, PS$\rightarrow$PL transfer, PL-side kernel execution, PL$\rightarrow$PS transfer, post-processing, and other minor overheads. Among them, Input Preparation is executed on the PS side and includes image loading, preprocessing, token generation, and quantization before FPGA execution, while PS$\rightarrow$PL and PL$\rightarrow$PS Transfer represent the data movement between the ARM processor and FPGA programmable logic. The PL Kernel stage denotes the execution latency of the event-sparse Linear Transformer accelerator, and Post-processing includes the final prediction decoding and classification operations. The Other category represents small control and synchronization overheads.As the sparsity configuration changes from K16/D16 to K11/D8, te PL Kernel latency decreases from 1146.1~$\mu$s to 868.4~$\mu$s, indicating that event-based sparsity can directly reduce the execution overhead of the FPGA computation core by reducing the number of effective tokens participating in computation and compressing the feature channel dimension. Meanwhile, the Total E2E latency decreases from 26272.0~$\mu$s to 19486.9~$\mu$s, showing that the end-to-end inference latency also decreases as the sparsity level increases. However, the variation in Input Preparation is also affected by host-side preprocessing and I/O, while the PL Kernel latency more directly reflects the hardware execution benefit of Event Sparse.
Specifically, the Total E2E latencies of K16/D16, K14/D13, K13/D10, and K11/D8 are 26272.0~$\mu$s, 21964.1~$\mu$s, 20295.5~$\mu$s, and 19486.9~$\mu$s, respectively. The PS$\rightarrow$PL Transfer and PL$\rightarrow$PS Transfer stages remain relatively stable, indicating that the main latency changes come from the input preparation and PL Kernel computation stages. These results confirm that event-based sparsity reduces the measured PL Kernel latency, while the overall end-to-end latency is also affected by the host-side Input Preparation stage.

In summary, These results demonstrate that the proposed method is not a simple output-level masking strategy, but transforms dynamic token-channel sparsity in Linear Transformer into practical FPGA execution optimization.The experimental results show that, while maintaining high detection performance, event-based sparsity can reduce FPGA resource usage and decrease both kernel latency and end-to-end inference latency on the PYNQ-Z2 platform.

\subsection{Case Study 3: Malware Detection Performance Analysis}

\begin{figure}[!ht]
    \centering
    \includegraphics[width=1\linewidth]{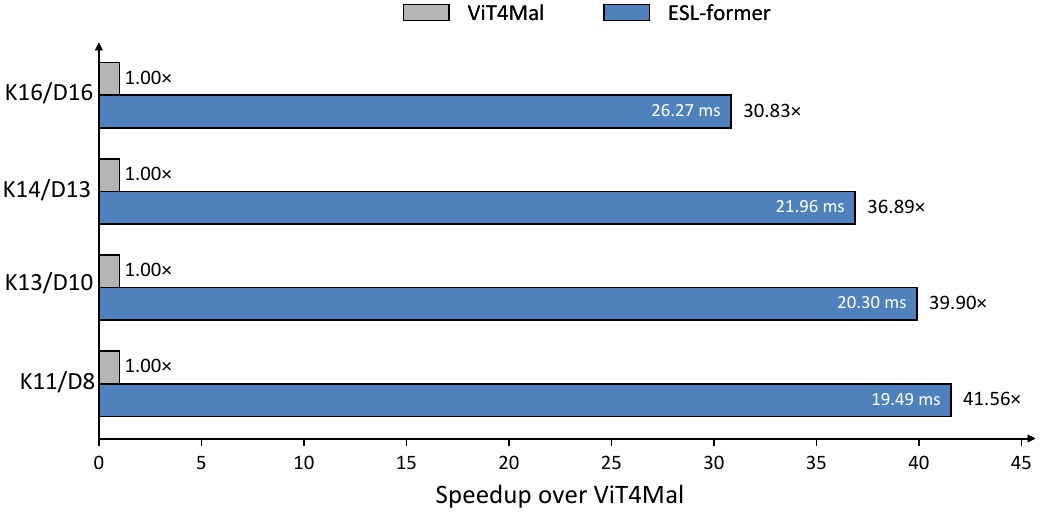}
    \caption{End-to-end latency reduction compared with ViT4Mal.}
    \label{fig:3}
\end{figure}

This experiment provides a system-level evaluation of the proposed method, including detection performance, FPGA resource utilization, end-to-end inference latency, and throughput. The results of ViT4Mal are taken from the published paper and are used as a system-level reference in this work.Table~\ref{tab:system_level_comparison} presents the system-level comparison between our method and representative lightweight malware image detection approaches, while Fig.~\ref{fig:3} further illustrates the end-to-end latency reduction of different Token-Channel sparse configurations compared with ViT4Mal. ViT4Mal is selected as representative baselines, and the proposed rawKV Event Sparse configurations with different Token-Channel settings are evaluated.

\begin{table}[!ht]
\centering
\caption{System-level comparison with existing malware detection methods on FPGA platforms.}
\label{tab:system_level_comparison}
\footnotesize
\setlength{\tabcolsep}{2.5pt}
\renewcommand{\arraystretch}{1.15}
\begin{tabular}{l*{8}{c}}
\toprule
Work &
\multicolumn{2}{c}{Detection} & \multicolumn{2}{c}{Lat./Thr.} & \multicolumn{4}{c}{FPGA Resource} \\
\cmidrule(lr){2-3} \cmidrule(lr){4-5} \cmidrule(lr){6-9}
 & Acc. & F1 & E2E & Thr. & LUT & DSP & FF & BRAM \\
\midrule
ViT4Mal~\cite{ravi2023vit4mal} & 92.41\% & -- & 810 ms & 1.23 & 16545 & 69 & 16911 & 92 \\
K16/D16 & 94.13\% & 96.53\% & 26.27 ms & 38.06 & 13701 & 149 & 12825 & 9.5 \\
K14/D13 & 93.35\% & 96.05\% & 21.96 ms & 45.53 & 11253 & 46 & 10588 & 5 \\
K13/D10 & 93.26\% & 95.99\% & 20.30 ms & 49.27 & 11161 & 46 & 10455 & 5 \\
K11/D8 & 92.78\% & 95.08\% & 19.49 ms & 51.32 & 10749 & 46 & 10299 & 5 \\
\bottomrule
\end{tabular}
\end{table}

From the detection performance, different Token-Channel sparse configurations maintain competitive detection capability. The K16/D16 configuration achieves an Accuracy of 94.13\% and an F1 score of 96.53\%. K14/D13 and K13/D10 achieve Accuracies of 93.35\% and 93.26\%, with F1 scores of 96.05\% and 95.99\%, respectively. When the compression level is further increased to K11/D8, the Accuracy and F1 decrease to 92.78\% and 95.08\%, respectively. These results indicate that Event Sparse can maintain stable detection performance under K16/D16, K14/D13, and K13/D10 configurations, while K11/D8 reflects the performance variation under a more aggressive sparsity level.

From the system latency and throughput, the end-to-end inference latency gradually decreases with increasing Token-Channel sparsity. The K16/D16, K14/D13, K13/D10, and K11/D8 configurations achieve end-to-end latencies of 26.27 ms, 21.96 ms, 20.30 ms, and 19.49 ms, corresponding to throughputs of 38.06 FPS, 45.53 FPS, 49.27 FPS, and 51.32 FPS, respectively. Compared with the 810 ms end-to-end latency of ViT4Mal, the proposed configurations achieve speedups of 30.83$\times$, 36.88$\times$, 39.90$\times$, and 41.56$\times$, as shown in Fig.~\ref{fig:3}.

From the perspective of FPGA resource utilization, different Token-Channel sparse configurations effectively reduce hardware resource requirements. The K16/D16 configuration uses 13,701 LUTs, 149 DSPs, 12,825 FFs, and 9.5 BRAMs. With further compression of the token number and channel dimension, the LUT usage of K14/D13, K13/D10, and K11/D8 decreases to 11,253, 11,161, and 10,749, while the FF usage decreases to 10,588, 10,455, and 10,299, respectively. Meanwhile, the DSP and BRAM usage remains relatively stable. Compared with ViT4Mal, the K11/D8 configuration achieves lower LUT, FF, and BRAM consumption while maintaining effective malware detection performance.

In summary, the K16/D16 configuration demonstrates that the Linear Transformer can effectively perform IMG\_DS malware image detection. The rawKV event-based sparse configurations further show that, while maintaining high detection performance, FPGA resource usage and end-to-end inference latency can be reduced. Comparisons against ViT4Mal further indicate that this work provides an effective hardware optimization approach for dynamic Token-Channel sparse execution of Linear Transformers on resource-constrained FPGA platforms.

%% file: sections/sec5_conclusion.tex
\section{Conclusion}\label{sec:conclusion}

This paper presented ESL-former, an event-sparse Linear Transformer framework for FPGA-accelerated malware image detection. The proposed method addresses two key limitations in deploying Transformer-based malware detectors on resource-constrained edge FPGAs.
Experimental results on the IMG\_DS malware image dataset and the PYNQ-Z2 platform demonstrate the effectiveness of the proposed approach. Compared with mask-based sparsity under the same token selection results, event sparsity achieves up to 1.20$\times$ measured kernel speedup on PYNQ-Z2, indicating that the latency gain comes from reduced redundant hardware execution overhead.
System-level comparisons further show that ESL-former is competitive for edge malware detection. Compared with the FPGA-based ViT4Mal baseline, the proposed K13/D10 configuration reduces end-to-end latency from 810 ms to 20.30 ms and BRAM usage from 92 to 5, while improving accuracy from 92.41\% to 93.26\%. As an edge deployment solution, ESL-former achieves up to 41.56$\times$ end-to-end latency speedup with the K11/D8 configuration while maintaining an accuracy of 92.78\%.
Future work will focus on streaming inference and evaluation on larger FPGA platforms and additional malware datasets.